\documentclass[preprint,aps,tightenlines,showpacs,nofootinbib]{revtex4}
\usepackage{latexsym}
\usepackage[dvips]{graphicx}
\newcommand{\beq}{\begin{eqnarray}}
\newcommand{\eeq}{\end{eqnarray}}
\newcommand{\eq}{eqnarray}

\newcommand{\al}{{\alpha}}
\newcommand{\be}{{\beta}}
\newcommand{\ci}{\cite}
\newcommand{\ga}{{\gamma}}

\newcommand{\ep}{{\epsilon}}

\newcommand{\de}{{\delta}}
\newcommand{\De}{\Delta}

\newcommand{\la}{{\lambda}}
\newcommand{\La}{{\Lambda}}
\newcommand{\m}{{\mu}}
\newcommand{\n}{{\nu}}

\newcommand{\Om}{{\Omega}}
\newcommand{\pa}{{\partial}}
\newcommand{\no}{{\nonumber}}
\newcommand{\f}{\frac}
\newcommand{\ra}{\rightarrow}
\newcommand{\lra}{\leftrightarrow}

\newcommand{\temp}{temperature }

\newcommand{\hO}{\hat{\Om}}

\newcommand{\na}{\nabla}
\begin{document}

\preprint{hep-th/0610140}

\title{Can Hawking temperatures be negative ?}

\author{Mu-In Park\footnote{E-mail address: muinpark@yahoo.com}}

\affiliation{ Center for Quantum Spacetime,  Sogang University,
Seoul 121-742, Korea \footnote{Present address: Research Institute
of Physics and Chemistry, Chonbuk National University, Chonju
561-756, Korea} }

\begin{abstract}
It has been widely believed that the Hawking temperature for a black
hole is $uniquely$ determined by its metric and $positive$. But, I
argue that this {\it might not} be true in the recently discovered
black holes which include the exotic black holes and the black holes
in the three-dimensional higher curvature gravities. I argue that
the Hawking temperatures, which are measured by the quantum fields
in thermal equilibrium with the black holes, might not be the usual
Hawking temperature but the $new$ temperatures that have been
proposed recently and can be $negative$. The associated new entropy
formulae, which are defined by the first and second laws of
thermodynamics, versus the black hole masses show some genuine
effects of the black holes which do not occur in the spin systems.
Some cosmological implications and physical origin of the
discrepancy with the standard analysis are noted also.
\end{abstract}


\maketitle

\newpage

\section{Introduction}

A black hole is defined by the existence of the non-singular event
horizon $r_+$, which is the boundary of the region of space-time
which particles or photons can escape to infinity, $classically$.
Bekenstein has shown that the black hole can be considered as a
``closed'' thermodynamical system with the temperature, proportional
to the surface gravity $\kappa_+$, and the chemical potentials,
proportional to the angular velocity $\Omega_{+}$ or electric
potential $\Phi_{+}$, if there is, at the horizon \cite{Beke:73}.
The argument was based on the Hawking's area (increasing) theorem
\cite{Hawk:71} and the black-hole analogue of the first law with the
temperature $T_+ \propto \kappa_{+}$, which is ``non-negative'', and
the entropy $S \propto {\cal A}_{+}$ for the horizon area ${\cal
A}_+$, which is ``non-decreasing'', i.e., satisfying the second law
of thermodynamics, due to the area theorem, as well as being
non-negative. Later, Hawking found that the black hole can radiate,
from the quantum mechanical effects, with the thermal temperature
$T_+=\hbar \kappa_{+}/2 \pi$ in accordance with the Bekenstein's
argument \cite{Hawk:75} [ I am using units in which $c=k_{B}=1$ ];
in this case, the black hole would not be a closed system anymore
but interacting with its environments such as the generalized second
law needs to be considered \cite{Hawk:75,Beke:74}.

There is an alternative approach to compute the Hawking temperature
by identifying $\hbar/T_{+}=2 \pi /\kappa_{+}$ as the periodicity of
the imaginary time coordinate which makes the metric regular at the
horizon \cite{Gibb:77} and this approach has been widely accepted;
no counter examples for this approach have been known so far, as far
as I know\footnote{Recently, it has been found that this approach
does not work anymore in the {\it smeared} black holes in the
quantum spacetime \ci{Kim:07}.}.
% \cite{Brow:91,Hawk:96}.
Now, since the surface gravity $\kappa_+$ at the horizon can be
computed from the metric unambiguously, the Hawking temperature in
this approach is $uniquely$ determined also. This would be the
origin of the widespread belief that the Hawking temperature $be$
uniquely determined by the metric in $any$ case. And also, it has
been widely believed that the Hawking temperature $be$ positive, as
in the Bekenstein's original argument \cite{Beke:73}. Actually, this
belief has been closely related to the ``positive mass theorems''
for black holes and the fact that the mass is grater than the
modulus of the charge, if there is \cite{Gibb:83}.

In this Letter, I argue that this belief {\it might not} be true in
the recently discovered black holes which include the exotic black
holes and the black holes in the three-dimensional higher curvature
gravities.

\section{New Hawking temperatures from thermodynamics}
\label{2}

In the spin systems the \temp can be negative, due to the upper
bound of the energy spectrum \cite{Kitt:67}. Recently, a number of
black hole solutions which have similar upper bounds of the black
hole masses have been discovered
\cite{Carl:91,Bana:98,Solo:06,Park:0602,Park:0609}. I have argued
that the Hawking temperatures for these systems might not be given
by the usual formula $T_+=\hbar \kappa_+/2 \pi$
\cite{Carl:91,Bana:98,Solo:06}, which is non-negative, but by new
formulae which $can$ be $negative$ depending on the situations
\cite{Park:0602,Park:0609}. The argument was based on the Hawking's
area theorem and the second law. This has been found to agree
completely with $CFT$ analysis, being related to the $AdS/CFT$
correspondence, as far as the $CFT$ analysis is available
\cite{Park:0602,Park:0609}. In this section let me briefly introduce
the black hole solutions and the thermodynamical arguments for the
new Hawking temperatures which differ from the usual formula and can
be negative.
%depending on the situations.

\subsection{The exotic BTZ black holes}

An exotic BTZ black hole is characterized by the following
properties

a. The metric is {\it formally} the same as the BTZ black hole
solution \cite{Carl:91,Solo:06,Park:0602,Park:0609}\footnote{But,
there is a qualitative difference from the BTZ solution for Einstein
gravity. This comes from the fact that the parameter $l$ appears
just as an integration constant in the black hole solution in Refs.
\cite{Carl:91,Solo:06,Park:0602}, though it is a parameter appearing
in the action for Einstein gravity. Actually, the suitable
distribution of matter replaces the cosmological constant in pure
gravity for the case of Ref. \cite{Carl:91}.}, which is given by
\cite{Bana:92},
\begin{eqnarray}
\label{BTZ}
 ds^2=-N^2 dt^2 +N^{-2} dr^2 +r^2 (d \phi +N^{\phi}
dt)^2
\end{eqnarray}
with
\begin{eqnarray}
N^2=\frac{(r^2-r_+^2) (r^2-r_-^2)}{l^2 r^2},~~ N^{\phi}=-\frac{r_+
r_-}{l r^2},
\end{eqnarray}
or modulus a 2-sphere \cite{Bana:98}. Here, $r_+$ and $r_-$ denote
the outer and inner horizons, respectively.

b. The mass and angular momentum, computed from the standard
Hamiltonian approach, are $completely$ interchanged from the
``bare'' ones $m$ and $j$ as
\begin{eqnarray}
\label{M_J}
 M=x j/l, ~~ J=x l m \label{MJ}
\end{eqnarray}
with an appropriate coefficient $x$; $x=1$ in Ref. \cite{Carl:91},
$x$ is a fixed value of $U(1)$ field strength in Ref.
\cite{Bana:98}, and $x$ is proportional to the coefficient of a
gravitational Chern-Simons term in Refs.
\cite{Solo:06,Park:0602}\footnote{It has been claimed that this
system, which goes beyond the physical bound to the coefficient of
the Chern-Simons term, is not well defined \ci{Solo:06}. But, the
argument has been based on the results which are valid only for
$|\be|/l <1$ and it does not apply to our case \ci{Park:0602}.}.
%,Park:06Com}.}.
Here, $m$ and $j$ are given by
\begin{eqnarray}
\label{bare:mj}
 m=\frac{r_+^2 +r_-^2}{8 Gl^2},~~j=\frac{2 r_+ r_-}{8Gl
},
\end{eqnarray}
which become the usual mass and angular momentum for the BTZ black
hole, with a cosmological constant $\Lambda=-1/l^2$, respectively
\cite{Bana:98}. The radii of the horizons are given by, in terms of
$m$ and $j$,
\begin{\eq}
\label{horizon}
 r_{\pm} =l \sqrt{ 4G m \left[1 \pm \sqrt{1-(j/ml)^2}\right]}
\end{\eq}
and it is clear, from this, that the bare parameters, which are
positive semi-definite, satisfy an inequality
\begin{\eq}
m \geq j/l
\end{\eq}
in order that the horizons exist ( the equality for the extremal
black hole with $r_+=r_-$ ). The remarkable result of (\ref{M_J}) is
that
\begin{eqnarray}
\label{M_bound}
 M^2 -J^2/l^2 =x^2 [j^2/l^2 -(m)^2] \leq 0
\end{eqnarray}
for $any$ non-vanishing $x$, which shows an upper bound for the mass
squared $M^2$ and a saturation for the extremal bare parameters,
i.e., $m=j/l$.

Now, given the Hawking temperature and angular velocity for the
event horizon $r_+$ of the metric (\ref{BTZ}), following the usual
approach \cite{Gibb:77},
\begin{eqnarray}
\label{T+}
 T_+=\left. \frac{\hbar \kappa}{2 \pi} \right|_{r_+}=\frac{\hbar
(r_+^2 -r_-^2)}{2 \pi l^2 r_+},~~\Omega_+=\left.-N^{\phi}
\right|_{r_+}=\frac{r_-}{l r_+}
\end{eqnarray}
with the surface gravity $\kappa=\partial N^2/2
\partial r$, the black
hole entropy has been identified as
\begin{eqnarray}
S=x \frac{2 \pi r_-}{4 G \hbar },
\end{eqnarray}
which satisfies the first law
\begin{eqnarray}
\label{first:old} \delta M=\Omega_+\delta J + T_+ \delta S
\end{eqnarray}
but depends on the $inner$ horizon area ${\cal A}_-=2 \pi r_-$
\cite{Carl:91,Bana:98,Solo:06}, rather than the outer horizon's
${\cal A}_+=2 \pi r_+$. But, there is $no$ physical justification
for this since the second law is not guaranteed
\cite{Park:0602,Park:0609} ( for an explicit demonstration, see Ref.
\cite{Park:0610} ). Rather, I have recently proposed another entropy
formula which does not have this problem
\begin{eqnarray}
S_{new}=|x| \frac{2 \pi r_+}{4 G \hbar}, \label{BH_new}
\end{eqnarray}
in accordance with the Bekenstein's original proposal
\cite{Beke:73}. Then, it is quite easy to see that this {\it does}
satisfy the second law since the metric (\ref{BTZ}) satisfies the
Einstein equation in vacuum, regardless of the details of the
actions,
\begin{\eq}
\label{eom}
 R^{\mu \nu}-\f{1}{2} g^{\m \n} R -\f{1}{{l}^2} g^{\m \n}
=0
\end{\eq}:
The Raychaudhuri's equation gives the Hawking's area theorem for the
outer horizon $\delta {\cal A}_+ \geq 0$, i.e., $\de S_{new} \geq 0$
since this vacuum equation satisfies the null energy condition
trivially \cite{Hawk:71,Hawk:75,Beke:74}; this can be also proved by
considering a ``quasi-stationary'' process which does $not$ depend
on the details of the gravity theory \cite{Jaco:95,Park:0610}. These
results are closely related to the fact that $dr_+/dm >0,~dr_-/dm
\leq 0$ for any (positive) $m$ and $j$ ( equality for $j=0$ ) since
these describe the rates of the area changes under the positive
energy matter accretion.

One interesting consequence of the new identification (\ref{BH_new})
is that I need to consider the rather unusual Hawking temperature
and angular velocity ( $\ep \equiv sign(x)$ )
\begin{eqnarray}
\label{T-}
 T_-=\ep \left. \frac{\hbar \kappa}{2 \pi} \right|_{r_-}=\ep
\frac{ \hbar (r_-^2 -r_+^2)}{2 \pi l^2
r_-},~~\Omega_-=\left.-N^{\phi} \right|_{r_-}=\frac{r_+}{l r_-},
\end{eqnarray}
respectively \cite{Park:0602,Park:0609}\footnote{This does not mean,
of course, that one needs an observer sitting on the inner horizon
$r_-$ to measure $T_-$ and $\Omega_-$, as it does not for
$T_+,~\Om_+$. }, such as the first law, as well as the manifest
second law, be satisfied also
\begin{eqnarray}
\label{SL:new}
 \delta M=\Omega_-\delta J + T_- \delta S_{new}.
\end{eqnarray}
Here, I note that, with these correct values of $M$, $J$, and the
entropy (\ref{BH_new}), which is proportional to the outer horizon
area, there is no other choice in the temperature and angular
velocity in the first law. The (positive) numerical coefficient in
the temperature $T_{-}$ of (\ref{T-}) is not determined from the
thermodynamical arguments but needs some other independent
identifications: This has been confirmed $indirectly$ in a $CFT$
analysis in Refs. \cite{Park:0602,Park:0609}; however, in this
Letter I support this, in a more traditional way, by identifying the
Hawking temperature $directly$ from the Green function analysis for
a quantum field. But, it is important to note that, regardless of
the numerical ambiguity, the temperature $T_-$ becomes ``negative''
$always$ for $x>0$. This can be easily understood from the existence
of the upper bound of mass $M \leq J/l$ with positive $M$ and $J$,
as in the spin systems \cite{Kitt:67}\footnote{The mass bound and
its resulting negative temperature might be related to the {\it
semiclassical} instability that has been found, recently
\ci{Bala:05}. Actually, if one applies the first and second laws as
in this Letter, the system of Ref. \ci{Bala:05} has a negative
temperature also due to to the {\it negative} mass, though not well
explored in detail in the literatures \ci{Cai:02}. So, I can suspect
that the negative temperature {\it might} be a signal of the same
instability as in Ref. \ci{Bala:05} and the negative temperature
spin systems in the ordinary surroundings with positive temperature
\ci{Kitt:67}. But the detailed analysis may be beyond the scope of
this Letter. I thank the referee for drawing my attention to this
interesting problem. }. Whereas, the temperature becomes positive
for $x<0$ due to the lack of an upper bound, i.e., $J/l \leq M$ with
$negative$ $M$ and $J$. These behaviors can be nicely captured in
the entropy, as a function of $M$ and $J$ (Fig.1), using (\ref{M_J})
and (\ref{horizon}):
\begin{eqnarray}
S_{new}=|x| \frac{2 \pi l}{4 G \hbar } \sqrt{ (4G J/x l) \left[1 +
\sqrt{1-(Ml/J)^2}\right]}.
 \label{BH_new:M}
\end{eqnarray}
\begin{figure}
\includegraphics[width=10cm,keepaspectratio]{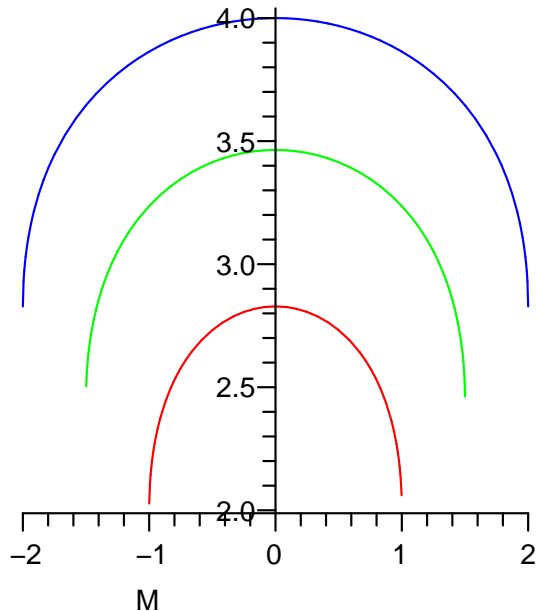}
\caption{The normalized entropies $S_{new} ( |x| 2 \pi l/4G
\hbar)^{-1}$ vs. $M$ for various values of $|J|/l$=1 (red), 1.5
(green), 2 (blue) [bottom to top] $(l=G=|x|=1)$.} \label{fig:exotic}
\end{figure}
Here, I note that the curves in Fig.1 are symmetric about $M=0$, as
in the spin systems: By the definition of the temperature
$1/T=(\partial S/ \pa M)_J$, I have $T_-<0$ on the right hand side
$(x>0)$, whereas $T_->0$ on the left hand side $(x<0)$; the two
temperatures $T_-=\pm \infty$ correspond to the same temperature for
a vacuum with $M=0$. But, note also that the entropy does $not$
vanish at the energy boundary $M=J/l$, i.e., extremal black hole,
and this would be inherent to black hole systems, which does not
occur in spin systems \cite{Wald:97}.

It is also important to note the fact, which is crucial in the
analysis of Sec. III, that the angular velocity has a lower bound
$\Omega_- \geq 1/l$, due to the fact of $r_+ \geq r_-$; it is
saturated by the extremal case $r_+=r_-$ and divergent in the limit
of $r_- \ra 0$. This implies that this system is always rotating, as
far as there is the event horizon $r_+$. And also, as $r_- \ra 0$,
this seems to be consistent with the fact of a non-vanishing angular
momentum $J$ since it satisfies the conventional relation $J \propto
\Omega M$, with the angular velocity $\Om=\Om_-$.
%On the other hand,
%for the usual choice of $\Om=\Om_+$, one has a nonsensical
%circumstance of a vanishing $\Om$, as well as $M$, as $r_- \ra 0$,
%though the angular momentum $J$ does not; as a result, the relation
%$J \propto \Omega M$ does not hold for $\Om=\Om_+$.
%Moreover, the
%choice of $\Om=\Om_-$ would be natural also in the sense that, if I
%consider the black hole as a ``rotating body'' whose angular
%velocity is determined by $(\partial M/\pa J)|_{r_+}$, regardless of
%the issue of the correct form of the entropy, its size should be
%characterized by its outer horizon $r_+$, not $r_-$ as is the case
%of $\Om_+$ in (\ref{first:old}).

\subsection{The BTZ black hole with higher curvatures}

The (2+1)-dimensional gravity with the higher curvature terms and a
`` bare '' cosmological constant $\Lambda=-1/{l}^2$ can be generally
described by the action [ omitting some boundary terms ]
\begin{\eq}
\label{Higher} I_{g}=\frac{1}{16 \pi G} \int d^3 x \sqrt{-g} \left(
~f(g^{\m \n}, R_{\m \n}, \nabla_{\m}) +\frac{2} {{l}^{2}}~\right),
\end{\eq}
where $f(g^{\m \n}, R_{\m \n}, \nabla_{\m})$ is an $arbitrary$
scalar function constructed from the metric $g^{\m \n}$, Ricci
tensor $R_{ \m \n}$, and the covariant derivatives $\nabla_{\m}$
\cite{Iyer:94,Said:00}. The equations of motion are
\begin{\eq}
\label{eom:Higher}
 \f{\pa f}{\pa g_{\mu \nu}}-\f{1}{2} g^{\m \n} f -\f{1}{{l}^2} g^{\m \n}
=t^{\m \n},
\end{\eq}
where $t^{\m \n}$ is given by
\begin{\eq}
t^{\m \n}=\f{1}{2} ( \na^\n \na^\al {P_\al}^\m + \na^\m \na^\al
{P_\al}^\n -\Box P^{\m \n}-g^{\m \n} \na^\al \na^\be P_{\al \be} )
\end{\eq}
with $P_{\al \be} \equiv g_{\al \m} g_{\be \n} (\pa f/\pa R_{\m
\n})$.

In the absence of the higher curvature terms, the BTZ solution
(\ref{BTZ}) is the $unique$ black hole solution in vacuum. Whereas,
even in the presence of the generic higher curvature terms, the BTZ
solution can be still a solution since the $local$ structure would
be ``unchanged'' by the higher curvatures: Actually $t_{\m \n}=0$
for the BTZ solution and the only effects are some
``re-normalization'' of the bare parameters $l,r_{\pm}$, and the
Newton's constant $G$, giving the Einstein equation
\begin{\eq}
\label{eom:ren}
 R^{\mu \nu}-\f{1}{2} g^{\m \n} R -\f{1}{l^2_{ren}} g^{\m \n} =0
\end{\eq}
in the renormalized frame \cite{Saho:06,Park:0609}. The renormalized
cosmological constant $\La_{ren}=-1/l^2_{ren}$ depends on the
details of the function $f$, but the renormalized Newton's
constant\footnote{Recently, this idea has been generalized to more
general class of black holes, including supergravity black holes
\ci{Brus:07}. } is given by
\begin{\eq}
G_{ren}=\hO^{-1} G
\end{\eq}
with
\begin{\eq} \label{Om}
 \hO \equiv \f{1}{3} g_{\m \n} \f{\pa f}{\pa R_{\m \n}},
\end{\eq}
which is constant for any constant-curvature solution
\cite{Said:00}. Now, due to the renormalization of the Newton's
constant, the original mass and angular momentum in (\ref{bare:mj})
are modified as
\begin{\eq}
\label{MJ:higher}
 M=\hO m,~~J=\hO j ,
\end{\eq}
respectively, by representing $m$ and $j$ as those in the
renormalized frame $m=\frac{r_+^2 +r_-^2}{8 G l^2_{ren}},~~j=\frac{2
r_+ r_-}{8G l_{ren} }$, with the renormalized parameters
$l_{ren},r_\pm$, but still with the bare Newton's constant $G$, such
as $m\geq j/l_{ren} $ is valid. Here, it is important to note that
$\hO$ is {\it not} positive definite \footnote{This means a negative
Newton's constant, but there is no a priori reason to fix the sign
in three dimensional gravities \cite{Dese:84}. This does not affect
the cosmic censorship condition in the Einstein frame either since
the (three-dimensional) frame transformation $\bar{g}_{\m
\n}=\hat{\Om}^2 g_{\m \n}$ is insensitive to the sign of $\hat{\Om}$
\ci{Jaco:95,Said:00}.
%On the other hand, the corresponding boundary
%CFT, which has negative central charge $c$ and conformal weight
%$L_0$, is known to be well defined also, by considering another
%representation of the Virasoro algebra with $\hat{L}_n =-L_{-n},
%\hat{c}=-c$, and the condition $\hat{L}_{n}|\hat{h}>=0~ (n>0)$ for
%the new height-weight state $|\hat{h}>$
%\ci{Bana:99,Park:0602,Park:0609}.
}, such as the usual inequality for the mass and angular momentum
would not be valid in general,
\begin{\eq}
\label{MJ:bound:higher}
 M-J/l= \hO (m-j/l)
\end{\eq}
but depends on the sign of $\hO$: $M \geq J/l$ for $\hO>0$, but $M
\leq J/l$ for $\hO <0$.

Regarding the black hole entropy, it has been computed as
\begin{eqnarray}
\label{S:Wald}
 S_W=\hO \frac{2 \pi r_+}{4 G \hbar}
\end{eqnarray}
from the Wald's entropy formula \cite{Iyer:94,Said:00,Saho:06}. But,
this is problematic for $\hO <0$, though it satisfies the usual
first law (\ref{first:old}), since $\de S_{W} \leq 0$ from the area
theorem which works in this case also due to the above Einstein
equation (\ref{eom:ren}) that satisfies the null energy condition in
the renormalized frame also, trivially. So, I have recently proposed
the modified entropy
\begin{eqnarray}
\label{S:Wald:mod}
 S_{W'}=|\hO| \frac{2 \pi r_+}{4 G \hbar},
\end{eqnarray}
which agrees with $CFT$ result as well \cite{Park:0602,Park:0609}.
Then, I need to consider the modified temperature ${T_+}'=sign(\hO)
~T_+$ in order to satisfy the first law
\begin{eqnarray}
\label{first:mod} \delta M=\Omega_+\delta J + {T_+}' \delta S_{W'}.
\end{eqnarray}
\begin{figure}
\includegraphics[width=10cm,keepaspectratio]{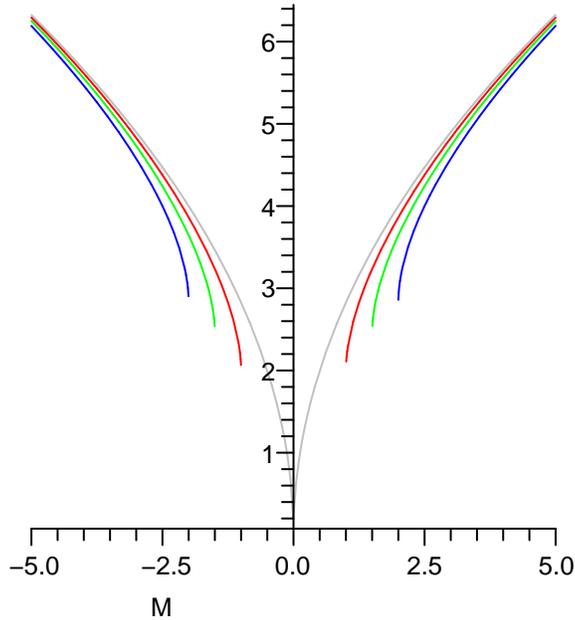}
\caption{The normalized entropies $S_{W'} ( |\hO| 2 \pi l_{ren}/4G
\hbar)^{-1}$ vs. $M$ for various values of $|J|/l$=0 (grey), 1
(red), 1.5 (green), 2 (blue) [top to bottom] $(l_{ren}=G=|\hO|=1)$.}
\label{fig:higher}
\end{figure}
The negative temperature ${T_+}'$ for $\hO <0$ is consistent with
the upper bound of mass $M \leq J/l$. The whole behaviors of the
temperature can be easily captured in the entropy, as a function of
$M$ and $J$ (Fig.2), using (\ref{horizon}) and (\ref{MJ:higher}):
\begin{eqnarray}
S_{W'}=|\hO| \frac{2 \pi l_{ren}}{4 G \hbar } \sqrt{ (4G M/\hO )
\left[1 + \sqrt{1-(J/Ml_{ren})^2}\right]}.
 \label{BH_new:M}
\end{eqnarray}
As can be observed in Fig.2, this system provides an unusual
realization of the negative temperature, which does not occur in the
usual spin systems: For $J\neq 0$, the two branches ($M>0$ and
$M<0$) are disjointed, due to a gap in the mass spectrum; the left
branch ($M<0$) has an upper bound $M<J/l$ and negative temperature
${T_+}'=-T_+$, whereas the right branch ($M>0$) has no upper bound
and so has the usual positive temperature $T_+$, following the usual
definition. According to the statistical mechanics, the gap is
natural because negative temperature is hotter than positive one
with the inequality $T=0_{+} <T=\infty_{+} <T=\infty_{-}<T=0_{-}$:
The left edge of $M>0$ curve, which has $T=0_{+}$, can not be
smoothly connected to the right edge of $M<0$ curve, which has
$T=0_{-}$; rather, the infinite right edge, which has
$T=\infty_{+}$, may be connected to the infinite left edge, which
has $T=\infty_{-}$. In this context, the $J=0$ cases whose curves
meet at $M=0$ does not
  seem to imply the {\it un-}bounded mass. Actually, the $M=0$ case can not be considered
  as the black hole spectrum because there is no horizon either;
  there is a discontinuity in the mass spectrum at
  $M=0$, which may be considered as an (open) upper bound for the left
  branch.

\section{Hawking temperatures from the Green functions}

The Hawking temperature can be fundamentally determined by the
periodicity of the thermal Green functions \cite{Hart:76}. In the
usual black hole systems this agrees with the periodicity for a
regular Euclideanized metric at the event horizon $r_+$. Actually,
the Hawking temperature for the BTZ metric has been determined in
this way and found to be the same as $T_+$ of (\ref{T+})
\cite{Lifs:94}. So, according to the widespread belief that Hawking
temperature be uniquely determined by the metric, the new Hawking
temperatures which do not agree with the usual temperature might be
considered as unphysical ones. But in this section I argue that this
{\it might not} be true in general, like as in the systems that I
have introduced in Sec. II: There were some ``loopholes'' in the
usual analyses which were unimportant for the ordinary black holes.
The results are consistent with the proposals of Sec. II.

To this end, I first note that the Hartle-Hawking Green function for
a scalar field in the background metric (\ref{BTZ}) is given by [ I
follow the approach of Ichinose-Satoh in Ref. \cite{Lifs:94}
]\footnote{For the system of Sec. IIB, the renormalized parameters,
$l_{ren},r_{\pm}$, are considered, instead.}
\begin{\eq}
- i G_{BH} (x,x')
    = \hbar (4 \pi l)^{-1} \sum_{n = - \infty}^{\infty} ( z_n^2 - 1)^{-1/2}
           [ z_n + ( z_n^2 - 1 )^{1/2} ]^{1-\lambda},
           \label{GBH}
\end{\eq}
where $x,x'$ are the points in the four dimensional embedding
space\footnote{The extra coordinates are frozen for the system of
Ref. \cite{Bana:98}. } and
\begin{\eq}
z_n(x,x') - i \varepsilon &=& d^{-2}_H
         \left[ \sqrt{r^2 -r_-^2} \sqrt{r'^2 - r_-^2}~
         \mbox{cosh} \left( {r_-}{l^{-2}} \Delta t - {r_+}{l^{-1}}
         \Delta \phi_n
               \right) \right.
               \no \\
      & & \left. \qquad \qquad -
          \sqrt{r^2 -r_+^2} \sqrt{r'^2 - r_+^2}
         ~\mbox{cosh} \left( {r_+}{l^{-2}} \Delta t - {r_-}{l^{-1}}
         \Delta \phi_n
               \right) \right]
               \label{zn}
\end{\eq}
with $d_H^2=r_+^2-r_-^2,~ \Delta t = t - t',~ \Delta \phi_n  = \phi
- \phi' + 2 n \pi $, and an infinitesimal positive imaginary part $i
\varepsilon$ [ the number $\la$ is a positive number which depends
on the scalar field's mass and its coupling to the metric
\cite{Lifs:94} ]. Here, it important to note that $z_n$, and so
$G_{BH}$, is symmetric under $r_+ \lra r_-$ interchange; this would
be a natural consequence of the symmetry in the metric (\ref{BTZ})
itself. Then, the Green function on the Euclidean black hole
geometry with the Euclidean time $\tau=it$ and the ``Euclidean''
angle $\varphi=-i \phi$ for $r_- \neq 0$ is
\begin{\eq}
G^{Eucl}_{BH} ( \Delta \tau, \Delta \varphi; r, r')=i G_{BH} (
\Delta t, \Delta \phi; r, r')\vert_{{\De \tau=i \De t}\atop{\De
\varphi = - i \De \phi}}.
\end{\eq}
The temperature, now, would be determined by comparing with the
$thermal$ Green function at temperature $\beta^{-1}$ and with a
chemical potential $\Om$ conjugate to angular momentum [ ${\cal T}$
denotes the Euclidean time ordered product for scalar fields
$\psi(x)$, and $\hat{H}$ and $\hat{J}$ are the generators of time
translation and rotation, respectively ],
\begin{\eq}
 G_{\beta}^{Eucl} (x, x'; \Om) = {\rm tr} \
          [ \ e^{ -\beta( \hat{H}- \Om \hat{J})} {\cal T}
          ( \psi(x) \psi(x') ) ]
          {/} {\rm tr} \
          [ \ e^{ -\beta( \hat{H}- \Om \hat{J})}],
\end{\eq}
which has the following periodicity:
\begin{\eq}
G_{\beta}^{Eucl} (\tau, \varphi, r; \tau', \varphi', r'; \Om) =
  G_{\beta}^{Eucl}
    (\tau + \beta \hbar , \varphi -  \Om \beta \hbar , r; \tau', \varphi', r';
    \Om).
  \label{period}
\end{\eq}
Because the Green function $G_{BH}$ is a function of $z_n$, one can
find, from (\ref{zn}), that $G^{Eucl}_{BH}$ is periodic under the
variation, with $({ m},{ n} \in {\bf Z})$,
\begin{\eq}
\de(\tau/l)= 2 \pi l d_H^{-2}(-r_- { m} + r_+ { n}),~ \de(\varphi)=
2 \pi l d_H^{-2}(r_+ {m} - r_- { n}).
\end{\eq}
If one requires that, as $r_- \ra 0$, the chemical potential $\Om$,
which being the angular velocity in a rotating black hole,
$vanishes$, the fundamental period is determined uniquely as
\begin{\eq}
\tau  \ra \tau + { 2 \pi }{\kappa_+^{-1} } { n},  ~~
  \varphi \ra \varphi - { 2 \pi  }{\kappa_+^{-1}\Om_+} { n}
\end{\eq}
with the angular velocity $\Om_+$ and the temperature
$\be^{-1}=\hbar \kappa_+/2 \pi$ as in (\ref{T+}); this is the usual
result \cite{Lifs:94}. But, this does not apply to the exotic
systems of Sec. IIA: The chemical potential $\Om_-$, which is
defined basically by the first law (\ref{SL:new}) or
(\ref{first:mod}) and also by the correct form of entropy
(\ref{BH_new}) or (\ref{S:Wald:mod}), respectively, which would
respect the second law, does $not$ vanish as $r_- \ra 0$ but
actually it has a ``lower'' bound $\Om_- \geq 1/l$ from (\ref{T-})
\cite{Park:0602}.
%So, in this case,
Now, for the generality, let me just assume the existence of the
lower bound only, regardless of its details. Then, it is easy to see
that the fundamental period may be determined ``uniquely'' as
\begin{\eq}
\tau  \ra \tau + { 2 \pi }{\kappa_-^{-1} } { m},  ~~
  \varphi \ra \varphi - { 2 \pi  }{\kappa_-^{-1}\Om_-} { m},
\end{\eq}
giving the angular velocity $\Om_-$ and the Hawking temperature
$\be^{-1}=\hbar \kappa_-/2 \pi$ as in (\ref{T-}), for $x>0$. For
$x<0$, on the other hand, the positive temperature $\be^{-1}=-\hbar
\kappa_-/2 \pi$ may be also determined
%uniquely
by considering
$(\hat{H},\hat{J},\beta) \ra (-\hat{H},-\hat{J},-\beta)$, in
accordance with the negative $M$ and $J$, from (\ref{M_J}). For the
system of Sec. IIB with $\hO <0$, in which the angular velocity
$\Om_+$ vanishes as $r_- \ra 0$ though, the temperature may be
%is uniquely
determined as $\be^{-1}=-\hbar \kappa_+/2 \pi$, which being
negative, with the ordinary angular velocity $\Om_+$ as in
(\ref{first:mod}), as well as the usual temperature $\be^{-1}=\hbar
\kappa_+/2 \pi$ for $\hO >0$. These results are consistent with the
proposals of Sec. II and agree completely with $CFT$ analyses
\cite{Park:0602,Park:0609}, as far as CFT analysis is available
\footnote{In the case of the BTZ black holes with a gravitational
Chern-Simons term, the boundary stress tensor is different from the
standard one, i.e., $\tau_{ij}=\ep^2 ({2 \be_{kl} }/{l^2
\sqrt{\ga^{(0)}}}) [ \ep^{jk} {\ga^{(2) i}}_{k} +(i \lra j) ]$ (see
Ref. \ci{Krau:06} for the details) and this produces the mass and
angular momentum (\ref{M_J}) with $x=32 \pi G \be_{KL}/l$ in the
standard definitions $M=l \oint d \phi \tau_{tt},~J=-l \oint d \phi
\tau_{t \phi}$. However, if we ``naively'' compute the corresponding
CFT operators, in the standard definition, one finds
$L_0^{\pm}-c^{\pm}/24=(l M \pm J)/2= \pm x (lm \pm j)/2$ with the
central charge $c^{\pm}= \pm x (3l/2G)$ and the conformal weight
$L_{0}^{\pm}$. This representation of the CFT can have negative
central charge $c$ and conformal weight $L_0$ but it is known to be
well defined also, by considering another representation of the
Virasoro algebra with $\hat{L}_n =-L_{-n}, \hat{c}=-c$, and the
condition $\hat{L}_{n}|\hat{h}>=0~ (n>0)$ for the new height-weight
state $|\hat{h}>$ \ci{Bana:99,Park:0602,Park:0609}. This twist is
not obvious in the stress tensor, but this is needed in order to
have a well-defined Hilbert space without the negative-norm states.
This looks to be also true in the higher curvature gravities though
the explicit form of the boundary stress tensor is not known
\ci{Said:00,Park:0602}. But, the corresponding explanations in other
cases are not available at present due to lack of the knowledge of
the boundary CFTs.}. These systems
%show
also suggest that the temperature
%is
{\it might not} be uniquely determined by the metric.
%, in contrast to the widespread belief.

\section{Concluding remarks}

So far, I have considered the cases which are described by the
three-dimensional metric (\ref{BTZ}), up to extra sphere parts. But,
there are also several other higher-dimensional black hole systems
which show negative Hawking temperatures, though not well recognized
in the literatures. The $AdS$ black holes in higher derivative
gravities \cite{Cvet:02} and the phantom (haired) black holes
\cite{Gao:06} are the examples ( see Ref. \cite{Park:0610} for the
details ). The implications of these black holes to the evolution of
the Universe filled with the phantom energy would be quite
interesting: If I consider the accretion of the phantom energy onto
a black hole with ``negative'' Hawking temperature, the black hole
size increase \cite{Diaz:04a}, as in the wormhole cases
\cite{Diaz:04b} but in contrast to the ordinary black holes with
positive Hawking temperatures \cite{Babi:04}, until a thermal
equilibrium with an equilibrium temperature is reached. This
equilibrium may be actually possible and can occur {\it before} the
catastrophic situations in Ref. \cite{Diaz:04b} if the phantom
energy has the negative temperature as claimed in Ref.
\cite{Diaz:04a}. Furthermore, the generalized second law of the
phantom Universe with a black hole can be satisfied also with the
negative Hawking temperature \cite{Diaz:04a}. The details will
appear elsewhere.

Finally, I would like to note some physical origin of the
discrepancy with the standard analysis as in Hawking's original work
\cite{Hawk:75} which yields the Hawking temperature $T_+$
independently of any details of the gravitational theory or
assumptions about the first law holding. To this end, I first note
that standard result is true only for the ``Riemanian'' or its
equivalent ``dual'' in the teleparallel gravity \cite{Blag:03}.
Otherwise, the particle's trajectory is ``not'' completely
determined by the metric only, and actually this seems to be the
case for the models of this paper. But, the details look different,
depending on the models. (i). For example, for the model of Ref.
\cite{Carl:91}, which is a teleparallel gravity with a vanishing
curvature ( and cosmological constant), the torsion is ``not totally
antisymmetric'' (in other words, the quantity called ``contortion''
exits) such as this is not equivalent to the ``Riemanian'' geometry
\cite{Hehl:76}. And, a non-vanishing angular velocity $\Omega_{-}$,
which was crucial in the analysis of Sec. III, as well as a
non-vanishing angular momentum $J$, as $r_{-} \ra 0$, would be the
result of the torsion, though the detailed relation is not explored
here. (ii). For the model of Refs. \cite{Solo:06,Park:0602}, a
particle (or particle-like solution) would also undergo the
mass/angular momentum interchange as in the black hole case
(\ref{M_J}) since one can ``not'' distinguish, basically, the black
hole solution from the point particle solution, though its explicit
form is not known, at the asymptotic infinity wherein the conserved
ADM mass and angular momentum are computed \cite{Carl:05}: It would
behave as a ``gravitational anyon'', similar to Deser's for the
asymptotically flat space \cite{Dese:90}. And also, it seems that
there is an intimate relation between the torsion (or contortion,
more exactly) in Ref. \cite{Carl:91} and the gravitational anyon in
Refs. \cite{Solo:06,Park:0602}, due to the relation between the
torsion and spin \cite{Park:92}. (iii). For the model of Ref.
\cite{Bana:98}, the explanation is not yet quite clear, but I
suspect that the constant U(1) flux on the 2-sphere would have a
crucial role in the non-standard behavior of Hawking radiation also.

As an alternative aspect of the peculiarity of Hawking radiation for
our models, I would like to note also that dynamical geometry
responds differently under the emission of Hawking radiation, as I
have emphasized in Ref. \cite{Park:0602}, recently. For example, the
emission of energy $\omega$ would reduce the black holes's mass $M$
from the conservation of energy, but this corresponds to the change
of the angular momentum $j$ in the ordinary BTZ black hole context,
due to the interchange of the roles of the mass and angular momentum
as in (\ref{M_J}). This is in sharp contrast to the case of ordinary
black hole. This seems to be also a key point to understand the
peculiar Hawking radiation in our system, and in this argument the
conservations of energy and angular momentum, which are not well
enforced in the standard computation, have a crucial role. In this
respect, the Parikh and Wilczek's approach \ci{Pari:00}, which
provides a direct derivation of Hawking radiation as a quantum
tunneling by considering the global conservation law naturally,
would be an appropriate framework to study the problem.

Finally, regarding the microscopic origin, it seems that the
existence of the negative temperature might imply the {\it spin
network model} of the quantum black holes \ci{Rove:95}, analogous to
the ordinary spin systems which can have negative temperature. But
it is not clear in that context how the negative temperature is
activated in our exotic examples but not in the ordinary black
holes.

\section*{Acknowledgments}

I would like to thank Gungwon Kang and Segrei Odintsov for useful
correspondences. This work was supported by the Science Research
Center Program of the Korea Science and Engineering Foundation
through the Center for Quantum Spacetime (CQUeST) of Sogang
University with grant number R11 - 2005- 021.

%%%%%%%%%% References %%%%%%%%%%%%%%%%%%%%%%%%%
\newcommand{\J}[4]{#1 {\bf #2} #3 (#4)}
\newcommand{\andJ}[3]{{\bf #1} (#2) #3}
\newcommand{\AP}{Ann. Phys. (N.Y.)}
\newcommand{\MPL}{Mod. Phys. Lett.}
\newcommand{\NP}{Nucl. Phys.}
\newcommand{\PL}{Phys. Lett.}
\newcommand{\PR}{Phys. Rev. D}
\newcommand{\PRL}{Phys. Rev. Lett.}
\newcommand{\PTP}{Prog. Theor. Phys.}
\newcommand{\hep}[1]{ hep-th/{#1}}
\newcommand{\hepp}[1]{ hep-ph/{#1}}
\newcommand{\hepg}[1]{ gr-qc/{#1}}
\newcommand{\bi}{ \bibitem}
%%%%%%%%%%%%%%%%%%%%%%%%%%%%%%%%%%%%%%%%%%%%%%%

\end{document}